\documentclass[prl,aps,twocolumn,superscriptaddress,amsmath,amssymb]{revtex4}
\usepackage{amsmath}
\usepackage[english]{babel}
\usepackage{graphicx}
\usepackage{bm}

\begin{document}

\title{Interference phenomena and long - range proximity effect
 in clean superconductor -- ferromagnet systems}

\author{A.~S.~Mel'nikov}
\affiliation{Institute for Physics of Microstructures, Russian
Academy of Sciences, 603950 Nizhny Novgorod, GSP-105, Russia}
\author{A.~V.~Samokhvalov}
 \affiliation{Institute for Physics of Microstructures, Russian Academy
of Sciences, 603950 Nizhny Novgorod, GSP-105, Russia}
\author{S.~M.~Kuznetsova}
\affiliation{Institute for Physics of Microstructures, Russian
Academy of Sciences, 603950 Nizhny Novgorod, GSP-105,
Russia}
\author{A.~I.~Buzdin} \affiliation{Institut Universitaire
de France and University Bordeaux, LOMA UMR-CNRS 5798, F-33405
Talence Cedex, France}
\date{\today}

\begin{abstract}
We study peculiarities of proximity effect in clean superconductor
-- ferromagnet structures caused by either spatial or momentum
dependence of the exchange field.
 Even a small modulation of the exchange field along the
 quasiparticle trajectories is shown to provide a long range contribution to the
supercurrent due to the specific interference of particle- and
hole- like wave functions. The momentum dependence of the exchange
field caused by the spin -- orbit interaction results in the long
-- range superconducting correlations even in the absence of
ferromagnetic domain structure and can explain the recent
experiments on ferromagnetic nanowires.
\end{abstract}

\maketitle

The exchange field $h$ in ferromagnetic (F) metals is well known
to destroy Cooper pairs resulting, thus, in a strong decay of
superconducting (S) correlations in the F material and suppression
of Josephson current in SFS junctions
(see Refs.~\onlinecite{Buzdin-rew,BergEfVolk-rew} for review).
Considering the quantum mechanics of quasiparticle excitations
this destructive effect of the exchange field can be viewed as a
consequence of a phase difference $\gamma\sim L/\xi_h=2Lh/\hbar
V_F$ gained between the electron- and hole- like parts of the
total wave function at the path of the length $L$. Both in the
clean and dirty limits the measurable quantities should be
calculated as superpositions of fast oscillating contributions
$e^{i\gamma}$ from different trajectories and, thus, rapidly
vanish with the increasing distance from the SF boundary.

This textbook physical picture appears to be in sharp contrast
with a number of recent experiments
\cite{Robinson,Khaire,Sosnin,Xiao,Giroud,Co wire} which point to
an anomalously large length of decay of superconducting
correlations inside the F metal. As we can judge from the
observation \cite{Co wire} of a noticeable supercurrent through a
Co nanowire, this decay length can be of the order of half a
micrometer which well exceeds typical coherence lengths in
ferromagnets both in the clean and dirty limits. In the dirty
limit such strong proximity effect can hardly be explained even
taking account of long--range triplet correlations
\cite{BergEfVolk-rew} induced by the exchange field inhomogeneity.

Naturally, the inhomogeneity of the field ${\bf h}$ caused by the
ferromagnetic domain structure can improve the conditions of
Cooper pair survival in the clean limit as well. To suppress the
destructive trajectory interference mentioned above the domain
structure should cancel the phase gain $\gamma$ for a certain
group of quasiparticle trajectories. A simple example of such
phase gain compensation can be realized in a clean junction
consisting of two F layers with opposite orientations of magnetic
moment \cite{blanter,pajovic}. On the other hand in the diffusive
limit this compensation effect vanishes \cite{crouzy}. Note, that
the exchange field inhomogeneity along the quasiclassical
trajectory experiencing multiple reflections from the ferromagnet
surface can appear even in the absence of the spatial domain
structure. Indeed, the exchange field acting on band electrons in
a solid with a finite spin -- orbit interaction should obviously
depend on the quasiparticle momentum \cite{shekhter}: ${\bf
h}={\bf h}({\bf k})$. The normal quasiparticle reflection is
accompanied, of course, by the change in the momentum direction,
and, thus, by the change in the exchange field. The momentum
dependent ${\bf h}$ field can strongly affect the phase gain
$\gamma$ along the trajectories even in the F sample prepared in a
single domain state (as it has been done in experiments with Co
nanowires \cite{Co wire}).

The goal of this paper is to show that in the clean limit there
exists a possibility to cancel the particle -- hole phase
difference for a large group of quasiclassical trajectories due to
either spatial or momentum dependence of the exchange field. Such
set of trajectories provides a long--range contribution to the
Josephson current through a ferromagnetic system which decays at
the length scale characteristic for a nonmagnetic metal. We
consider two generic examples which illustrate the above scenario
of a long--range proximity effect: (i) Josephson transport through
a pair of ferromagnetic layers with a stepwise exchange field
distribution; (ii) Josephson transport through a nanowire with a
specular electron reflection at the surface and exchange field
varying with the changing quasiparticle momentum. See Supplemental
Material at [URL will be inserted by publisher] for details of
calculations.


\noindent{\it Josephson transport through a ferromagnetic
bilayer.}
%
\begin{figure}[htb!]
    \includegraphics[width=7.5cm]{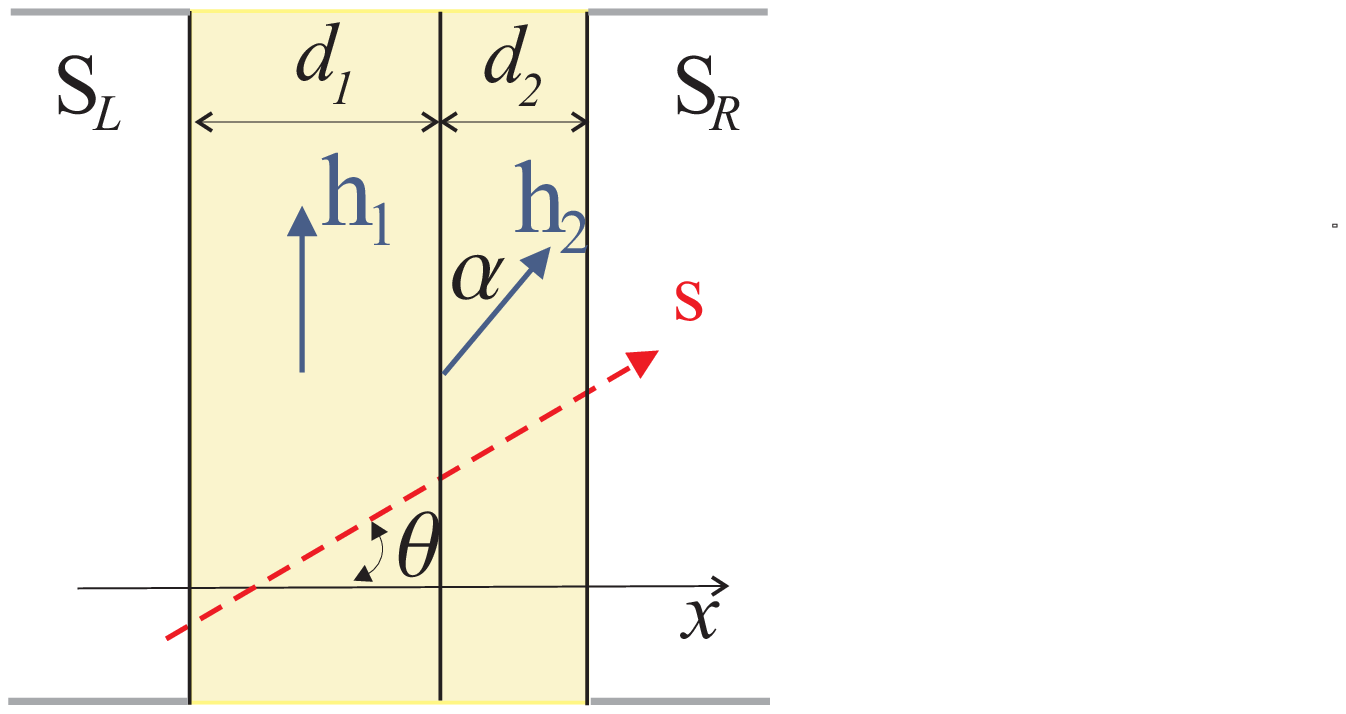}
    \caption{(Color online) Josephson junction containing two ferromagnetic layers.
  Linear quasiparticle trajectory is shown by the red dashed line.}
    \label{fig1}
\end{figure}
%
Let us start from the simplest model illustrating the
origin of the quasiparticle interference suppression: Josephson
junction containing two ferromagnetic layers of the thicknesses
$d_1$ and $d_2$, respectively (see Fig.1). Here we consider the limit of short
junction $d_1+d_2\ll\xi_s$, where $\xi_s$ is the superconducting
coherence length. The exchange fields ${\bf h}_1$ and ${\bf h}_2$
in the layers are rotated at the angle $\alpha$.
Following the quasiclassical procedure considered in
Ref.~\onlinecite{Buzdin-prb11} we
find the current -- phase relation:
\begin{equation}
I=\sum\limits_n I_n =\sum\limits_n a_n \sin n\varphi \frac{\langle
({\bf n},{\bf n}_F) \cos n\gamma\rangle}{\langle ({\bf n},{\bf
n}_F) \rangle} \ ,
\end{equation}
where ${\bf n}$ is the unit vector normal to the junction plane,
${\bf n}_F$ is the unit vector along the trajectory, and
 $a_n$ are the coefficients of the Fourier expansion for the
current -- phase relation $I_{SNS}(\varphi)$ for zero exchange
field, i.e., for superconductor -- normal metal junction of the
same geometry. The angular brackets denote the averaging over different
quasiclassical trajectories. The first two coefficients in this expansion take
the form:
%
\begin{equation}
    a_n=\frac{4 e T}{\hbar } N (-1)^{n-1}\sum_{m=0}^{\infty}
        \left( \mu_m - \sqrt{\mu_m^2-1} \right)^n, \; n=1,2,           \label{S3} \\
\end{equation}
where $\mu_m=2\pi^2 T^2 (2m+1)^2 / \Delta_0^2+1$, $\Delta_0$ is
the temperature dependent superconducting gap, $ N= s_0^{-1}\int
ds \int d{\bf n}_F (\mathbf{n}_F,\mathbf{n}) $,
$s_0^{-1}=k_F/2\pi$ ($s_0^{-1}=(k_F/2\pi)^2$) for 2D (3D)
junctions, and the integral $\int \ldots ds$ is taken over the
junction cross--section. The factor $N$ is determined by the
number of transverse modes in the junction: $N\sim S/s_0$, where
$S$ is the junction cross--section area.

%
The phase $\gamma$ can be found from the singlet part of the
anomalous quasiclassical Green function: %
$$f_s (s=s_R)=\cos\gamma$$
taken at the right superconducting
electrode. Here we use a standard parametrization \cite{Champel}
$f=f_s+\textbf{f}_t\hat\sigma$, where $\hat\sigma$ is a Pauli
matrix vector in the spin space. The functions $f_s$,
$\textbf{f}_t$ satisfy the linearized Eilenberger equations
written for zero Matsubara frequencies
\begin{equation}
-i\hbar V_F \partial_s f_s
+2\textbf{h}\textbf{f}_t=0 \ , \quad 
-i\hbar V_F \partial_s \textbf{f}_t
+2f_s\textbf{h}=0 \ , \label{eil-2}
\end{equation}
and the conditions $f_s(s=s_L)=1$, $\textbf{f}_t(s=s_L) =0$ at the
left superconducting electrode. Solving the above equations for
the particular bilayer geometry we find:
\begin{equation}
\cos\gamma
=\cos^2\frac{\alpha}{2}\, \cos\left(\frac{d_1+d_2}{\xi_h\cos\theta}\right)+
\sin^2\frac{\alpha}{2}\,\cos\left(\frac{d_1-d_2}{\xi_h\cos\theta}\right) \ ,
\end{equation}
where $\cos\theta = ({\bf n},{\bf n}_F)$. This expression allows
us to write the first harmonic in the current -- phase relation in
the form:
\begin{equation}
I_1= \left[\cos^2\frac{\alpha}{2}\,
I_{c1}\left(\frac{d_1+d_2}{\xi_h}\right)+\sin^2\frac{\alpha}{2}\,
I_{c1}\left(\frac{d_1-d_2}{\xi_h}\right)\right] \sin\varphi \ ,
\end{equation}
where $I_{c1}(d/\xi_h)$ is the critical current of the first
harmonic in a SFS junction with a homogeneous exchange field $h$.
The interference effects discussed in introduction result in the
power decay of the critical current $I_{c1}$ vs the F layer
thickness $d$: $I_{c1}\propto d^{-1/2}$ for a 2D junction
\cite{2D} and $I_{c1}\propto d^{-1}$ for a 3D junction \cite{3D}.
Taking symmetric case $d_1=d_2$ we immediately get a long--range
contribution to the Josephson current
\begin{equation}\label{eq:7}
\delta I_{c1}= \sin^2\frac{\alpha}{2}\,I_{c1}\left(0\right) \sin\varphi \ ,
\end{equation}
which does not decay with the increasing distance between the S
electrodes. It is important to note that this contribution {\it
does not vanish for an arbitrary nonzero angle} between the
magnetic moments in the F layers.

Long--range behavior can be observed for a second harmonic in
the current -- phase relation as well. Indeed, calculating the
average $\langle ({\bf n},{\bf n}_F) \cos 2\gamma\rangle$ we find
a nonvanishing long--range supercurrent contribution even for
$d_1\neq d_2$:
\begin{equation}
\delta I_{c2}= \frac{ a_{2}\sin^2\alpha}{2}\sin 2\varphi \ .
\end{equation}
Note, that the emergence of long--range proximity effect for
high harmonics in Josephson relation is in a good agreement with
recent theoretical findings in
Refs.~\onlinecite{trifunovic1,trifunovic2}.

\noindent{\it Josephson current through a ferromagnetic wire.} We
now proceed with the consideration of a more complicated example
of the interference phase suppression in a ferromagnetic wire
where the quasiclassical trajectories of electrons and holes
experience multiple specular reflections from the wire surface
(see Fig.~2a). The particular geometry shown in Fig.~2a can be
considered as a rough model for experiments on Co nanowires
\cite{Co wire}. For simplicity we restrict ourselves to the case
of a 2D junction.
\begin{figure}[htb!]
    \includegraphics[width=9cm]{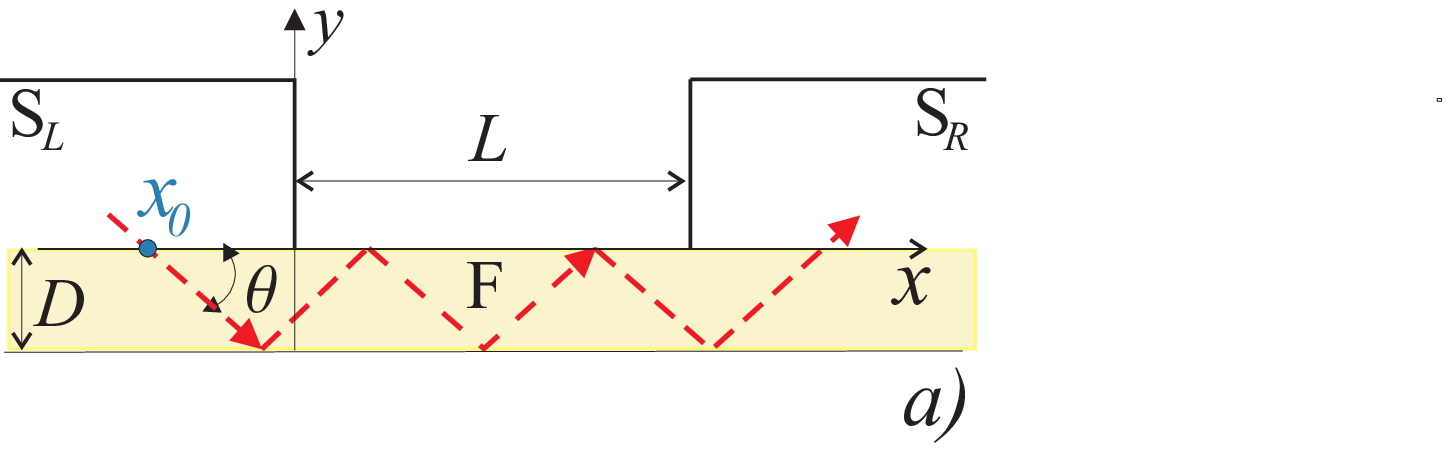}
    \includegraphics[width=9cm]{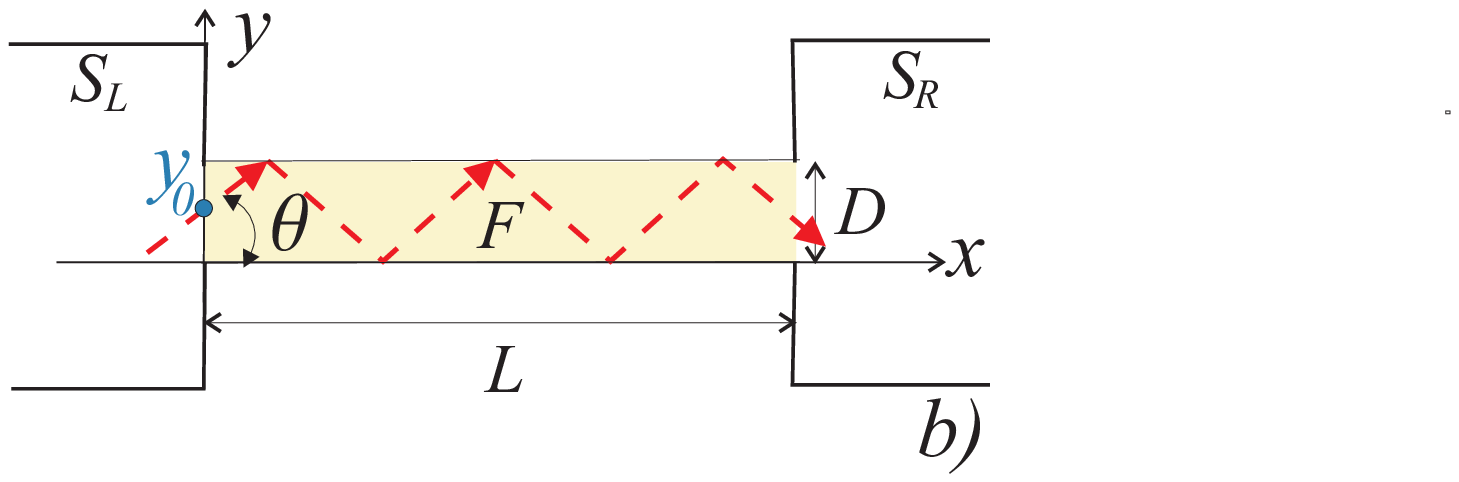}
    \caption{(Color online) Josephson transport through a nanowire
        in the overlap (a) and edge (b) geometries. The quasiparticle trajectories
        are shown by the red dashed lines.}
    \label{fig2}
\end{figure}

Taking account of the spin-orbit interaction inside the ferromagnet we obtaine
the exchange part of the effective Hamiltonian for the band electrons
depending on the quasi-momentum ($\mathbf{k}$) orientation:
$$ \hat H_{ex} = \sum\limits_{ij} \beta_{ij}(\mathbf{k}) h_{0i} \sigma_j =
\mathbf{h}(\mathbf{k})\hat\sigma\,,
$$
where $\mathbf{h}_0$ is a
pseudo vector determined by the ferromagnetic moment. Assuming the
absence of the system anisotropy described by a polar vector  we
find the simplest form of the resulting exchange field:
$\mathbf{h} = \mathbf{h}_0 + \beta_{so} k_F^{-2}
(\mathbf{h}_{0},\mathbf{k})\mathbf{k}$, where $\beta_{so}$ is a
constant determined by the spin -- orbit interaction, and $k_F$ is
the Fermi momentum.

The exchange field along the quasiparticle trajectory experiencing
the reflection at the wire surface should change its direction.
Thus, we obtain the problem described by Eqs.~(\ref{eil-2})
with a periodic exchange field along the trajectory
characterized by a given angle $\theta$ and a certain starting
point at the superconductor surface. The same equations for each
trajectory can be of course derived for a periodic domain
structure. Let us consider first the problem of calculating the
band spectrum $\epsilon (k)$ in the field $\mathbf{h}$ varying
with the period $2D/\sin\theta$:
\begin{eqnarray}
&&-i\hbar V_F \partial_s f_s
+2\textbf{h}\textbf{f}_t=\epsilon (k) f_s\,, \label{eps-1} \\
&&-i\hbar V_F \partial_s \textbf{f}_t
+2f_s\textbf{h}=\epsilon (k) \textbf{f}_t\,. \label{eps-2}
\end{eqnarray}
The solution can be written in the Bloch form:
$$
\left(f_s\atop\textbf{f}_t\right) =
e^{iks}\left(f_{sk}\atop\textbf{f}_{tk}\right) \ ,
$$
where $f_{sk}(s+2D/\sin\theta)=f_{sk}(s)$ and
$\textbf{f}_{tk}(s+2D/\sin\theta)=\textbf{f}_{tk}(s)$. One can see
that provided this solution corresponds to the energy branch
$\epsilon_\sigma (k)$ there exist another solution $(f_s^*,
-\textbf{f}_t^*)$ corresponding to the energy
$-\epsilon_\sigma(k)$. On the other hand the latter solution
corresponds also to the energy $\epsilon_{\tilde\sigma}(-k)$ and,
thus, we obtain the following symmetry property of the band
spectrum: $\epsilon_{\tilde\sigma}(-k)=-\epsilon_{\sigma}(k)$,
where the indices $\sigma$ and $\tilde\sigma$ denote different
branch numbers. The full set of energy branches can be split in
such pairs provided the number of branches is even. For an odd
number of branches there is always one branch which does not have
a partner. For this branch we obtain
$\epsilon_{\sigma}(-k)=-\epsilon_{\sigma}(k)$ and, thus, this
spectrum branch crosses the zero energy level at $k=0$:
$\epsilon_{\sigma}(0)=0$. The corresponding phase gain $\gamma$
appears to vanish for trajectories containing an integer number of
periods shown in Fig.~2a and, therefore, the solution with $k=0$
and $\epsilon=0$ provides a long--range contribution
to the supercurrent.

For the sake of definiteness we choose the field $\mathbf{h}_0$ to
be directed along the wire axis $x$ and obtain the exchange field
in the form: $ \mathbf{h} = \mathbf{x}_0 h_x + \mathbf{y}_0 h_y
(s)$, where $h_x(\theta) \simeq h_0$ is constant along the
trajectory and $h_y (s)$ is a
 periodic function with zero average. In the interval $-D/\sin\theta<s<D/\sin\theta$
 the $h_y$ field component is defined by the expression
$h_y=\beta_{so} h_0 \sin\theta\cos\theta\, {\rm sign}\,s $.
Introducing the Fourier expansions
\begin{eqnarray}
h_y=\sum\limits_q H_q \mathrm{e}^{iqs} \ ,\,\,\,\,\, H_q=-i\tilde
h\frac{2\sin\theta}{Dq}\,, \nonumber \\
f_{s,tx,ty} = \mathrm{e}^{iks}\sum\limits_q F_{s,x,y}(k+q) \mathrm{e}^{iqs}\,, \nonumber
\end{eqnarray}
we rewrite the Eqs.~(\ref{eps-1}) and (\ref{eps-2}) in the form:
\begin{eqnarray}
&&(\hbar V_F(k+q)-\epsilon)F_s(k+q)+2h_xF_x(k+q) \nonumber \\
&&\qquad\quad + 2\sum\limits_{\tilde q} H_{q-\tilde q} F_y(k+\tilde q)=0\,, \\
&&(\hbar V_F(k+q)-\epsilon)F_x(k+q)+2h_xF_s(k+q)=0\,,\qquad   \\
&&(\hbar V_F(k+q)-\epsilon)F_y(k+q)              \nonumber \\
&&\qquad\quad + 2\sum\limits_{\tilde q} H_{q-\tilde q}F_s(k+\tilde
q)=0\,.
\end{eqnarray}
Here $q,\tilde q=q_m=\pi (2m+1)\sin\theta/D$, $m$ is an integer,
and $\tilde h=\beta_{so} h_0 \sin\theta\cos\theta$.

To get the solution for a small periodic field $h_y$ we use a
perturbative approach similar to the nearly free electron
approximation in the band theory of solids and restrict the number
of interacting Fourier harmonics in the expansions. For this
purpose it is instructive to consider the limit of zero periodic
potential $h_y$ and separate three types of solutions: (i) the
solution $(F_s,F_x,F_y)=(0,0,1)\delta_{q-p}$ corresponding to the
energy $\epsilon_0=\hbar V_F (k+p)$ (ii) the solutions
 $(F_s,F_x,F_y)=(1, \pm 1,0)\delta_{q-p_\pm}$
corresponding to the energies $\epsilon_\pm=\hbar V_F (k+p_\pm)\pm
2h_x$. Here $p$ and $p_\pm$ are arbitrary reciprocal lattice
vectors. The above modes should strongly interact provided the
resonant condition $\epsilon_0=\epsilon_+=\epsilon_-$ is
fulfilled. Such resonance is possible for the case when the value
$2h_x/\hbar V_F$ equals to a certain reciprocal lattice vector
$q_m$. Close to such Bragg -- type resonance we see that the
dominant harmonics correspond to the following choice of
reciprocal lattice vectors: $p=0$, $p_\pm=\mp q_m$. Writing the
solution as a superposition of these three harmonics we find
renormalized spectral branches $\epsilon_0=\hbar V_F k$,
$\epsilon_{\pm}=\hbar V_F k\pm\sqrt{(\hbar
V_Fq_m-2h_x)^2+8|H_{q_m}|^2}$
 and corresponding eigenfunctions.
Applying now the boundary conditions at $s=0$ for the
superposition of the above eigenfunctions we find the amplitude of
the singlet component corresponding to the energy branch
$\epsilon_0$ and $k=0$:
$$ f_{sm} = \frac{8|H_{q_m}|^2  \cos(q_m s)}{(\hbar
V_Fq_m-2h_x)^2+8|H_{q_m}|^2} \ . $$
At the surface of a right superconducting electrode we should take
the coordinate $s$ to be equal to the integer number of periods.
We also need to sum up the above resonant expressions over all
Fourier harmonics of the periodic potential:
$$
f_s (s=s_{R}) = \sum\limits_{m=0}^\infty\frac{8|H_{q_m}|^2}{(\hbar
V_Fq_m-2h_x)^2+8|H_{q_m}|^2}  \ .
$$
The precision of such resonant -- type expression has been also
confirmed by the numerical solution of the Eqs.~(\ref{eps-1}) and
(\ref{eps-2}) carried out using the transfer matrix method. Note,
that we omit here the contribution from the solutions
corresponding to the branches $\epsilon_\pm$: these functions
correspond to a nonzero quasimomentum and, thus, should gain a
finite phase factor along the trajectory length. During averaging
over different trajectories this phase factor causes the
suppression of the resulting supercurrent contribution with the
increasing wire length $L$.

The starting point of the trajectory varies in the interval
$\Delta x= 2D/\tan\theta$ and, as a consequence, the long -- range
first harmonic in current -- phase relation takes the form:
$$ I_1 = a_1 \sin\varphi \int\limits_0^{\pi/2}d\theta\cos\theta
f_s (s_{R})\, . $$
Assuming the resonances to be rather narrow we approximate them by
the delta -- functions and obtain:
$$ I_1 = a_1 \sin\varphi \sum\limits_{m} \frac{\sqrt{2}\pi\hbar
V_F\tilde h (\theta_m)}{h_x^2D}\sin^2\theta_m\, . $$
where $\sin \theta_m = 2h_x D/\pi\hbar V_F(2m+1)$.
In the limit $D\gg \hbar V_F/2h_x$ one can replace the sum
over $m$ by the integral:
$$
I_1\simeq a_1\sqrt{2}\int\limits_0^{\pi/2} d\theta\frac{\tilde h
(\theta)}{h_x(\theta)}\cos\theta\sin\varphi \simeq
a_1\frac{\sqrt{2}}{3} \beta_{s0} \sin\varphi  \ .
$$

Certainly, the above long--range effect in the first harmonic is
rather sensitive to the system geometry: taking, e.g., the system
sketched in Fig.~2b we will not obtain the full cancellation of
the phase $\gamma$ because the trajectories in this case do not
contain integer number of exchange field modulation periods.
However, similarly to the case of bilayer the long--range effect
is still possible for higher harmonics. We apply the above
perturbative procedure for the calculation of the full $f_s$
function for the geometry shown in Fig.~2b.
%
%
%
The second harmonic in the
current--phase relation reads
\begin{equation}\label{eqn:I2}
I_2 = a_2 \sin 2\varphi\ \int\limits_0^{\pi/2}d\theta \cos\theta
\left(\, 2 \langle f_s^2 (s_{R}) \rangle_{y_0} - 1\, \right)\,,
\end{equation}
where $\langle \ldots \rangle_{y_0}=(1/D)\int_0^{D} \ldots dy_0$
denotes averaging over the starting point of the trajectory $y_0$
(see Fig.~2b). Keeping only the terms linear in the small
$|H_{q_m}|$ amplitude we get the following expression for the
long--range part of the second harmonic $I_2$:
$$ I_2 = a_2 \sin 2\varphi \sum\limits_{m} \frac{\sqrt{2}\pi\hbar
V_F\tilde h (\theta_m)}{h_x^2D}\sin^2\theta_m \simeq
a_2\frac{\sqrt{2}}{3} \beta_{s0} \sin 2\varphi . $$
We emphasize that the second harmonic of Josephson current in both
above examples is negative because of the condition $a_2<0$.

Note that the absence of the decay of the single-channel critical
current was pointed out in Ref.~\onlinecite{buzdin} as a possible
source of the long-ranged proximity effect in Co nanowires.
However the averaging of the phase gain for different modes
strongly decreases the critical current. In contrast the results
presented in this Letter demonstrate that in the ballistic regime
the spin-orbit interaction generates the non-collinear exchange
field which produces the long -- range Josephson current. This
conclusion is always true for the second harmonic in the current
-- phase relation and for some geometries it may be also valid for
the first harmonic.
Therefore our findings provide a natural explanation of the recent
experiments with Co nanowire \cite{Co wire}.
To discriminate between two proposed mechanisms of the long ranged effect,
the studies of higher harmonics in Josephson current-phase relations
could be of major importance.
Also it should be interesting to verify on experiment the predicted
simple angular dependence (\ref{eq:7}) of the critical current in S/F/S
junctions with composite interlayer.

\acknowledgments{ The authors thank R. Shekhter for valuable
comments. This work was supported, in part, by European IRSES
program SIMTECH (contract n.246937), the Russian Foundation for
Basic Research, FTP "Scientific and educational personnel of
innovative Russia in 2009-2013", and the program of LEA "Physique
Theorique et Matiere Condensee".}

\onecolumngrid

%
%
%

\end{document}